\documentclass{article}
\usepackage[utf8]{inputenc}
\usepackage{authblk}
\usepackage{setspace}
\usepackage[margin=1.45in]{geometry}
\usepackage{graphicx}
\usepackage{float}
\usepackage{changepage} 
\graphicspath{ {./figures/} }
\usepackage{subcaption}
\usepackage{amsmath}
\usepackage[hidelinks]{hyperref}

\usepackage[right]{lineno}
\usepackage{xcolor} 



\title{Review on recent results of J/$\psi$ production at STAR}

\author{Jitka Mrázková\textsuperscript{1,2,*} for the STAR Collaboration}

\date{}

\begin{document}
\maketitle
\vspace{-1cm}
\begin{center}
    {\small
    \textsuperscript{1}\itshape Faculty of Nuclear Sciences and Physical Engineering, Czech Technical University in Prague\\
    \textsuperscript{2}\itshape Nuclear Physics Institute, Czech Academy of Sciences\\
    \textsuperscript{*}Address correspondence to: \texttt{jitka.mrazkova@fjfi.cvut.cz}
    }
\end{center}
\vspace{0.5cm}

\onehalfspacing
\begin{abstract}
Studying the production of J/$\psi$ (bound state of charm and anti-charm quark) in proton-proton collisions gives an opportunity to test quantum chromodynamics (QCD) calculations, as the production of J/$\psi$
involves both perturbative and non-perturbative processes. However, theoretical calculations are still unable to fully and simultaneously explain experimental results, such as polarization and $p_\text{T}$ spectra measured in different kinematic regimes and colliding energies. More studies
are needed to investigate J/$\psi$ production mechanism. In heavy-ion collisions, charmonia can be used to study the properties of the medium as they are expected to dissociate in the medium when
the Debye radius, inversely proportional to the medium temperature, becomes smaller than their size. Other competing effects, such as recombination, have also been found to modify the
observed J/$\psi$ yield in heavy-ion collisions. We review recent measurements of the J/$\psi$ production in proton-proton and heavy-ion collisions at various collision energies measured with the STAR
experiment at RHIC. The data are compared with recent model calculations on charmonia production.
\end{abstract}


\vspace{0.5cm}

\section{Introduction}

Understanding the production mechanisms of J/$\psi$ mesons is crucial for testing
QCD calculations in proton-proton ($p+p$) collisions and probing the properties
of the strongly interacting medium created in heavy-ion collisions.
J/$\psi$ suppression provides evidence of QGP formation, where
color screening prevents the binding of charm quarks, depending
on the energy density and temperature of the medium \cite{quarkonia}. Describing charmonium production in a medium is challenging due to the competing effects of recombination and dissociation processes.
Systematic studies of various collision systems and energies
may help to disentangle the charmonium production mechanism. Accordingly, the following proceedings provide an overview of recent results on the J/$\psi$ production from the STAR experiment.
\section{Experimental setup}

All of the analyses presented here were obtained from data collected using the STAR (Solenoidal Tracking at RHIC) detector \cite{star}. Some of the key detectors utilized in J/$\psi$-related analyses at mid-rapidity include the Time Projection Chamber (TPC) \cite{tpc}, which provides charged particle tracking, momentum determination, and energy loss measurements for particle identification. The Time-of-Flight (TOF) detector \cite{tof} is used for complementary particle identification. The Barrel Electromagnetic Calorimeter (BEMC) \cite{bemc} operates as for particle detection based on deposited energy, provides high-$p_\text{T}$ triggering, and features fine granularity in  $(\eta,\varphi) = (0.05,0.05)$ with full azimuthal coverage: $0 \le \varphi <  2\pi$.

\vspace{1cm}

\section{Results}

Firstly, the results focused on suppression of the J/$\psi$ production, reconstructed via the dielectron decay channel, in heavy-ion collisions across various collision systems and energies will be presented. This will be followed by a discussion of the analyses related to the J/$\psi$ production in proton-proton collisions, including studies of multiplicity dependence and the J/$\psi$ production within jets.
\vspace{0.5cm}

\subsection{Inclusive J/$\psi$ R$_\text{AA}$}
To study the influence of the medium on the J/$\psi$ production, one can use the nuclear modification factor R$_\text{AA}$ defined as:
\vspace{0.5cm}
\begin{equation}
    R_{A A}=\frac{\sigma_{\text {inel }}}{\left\langle N_{\text {coll }}\right\rangle} \frac{\text{d}^{2} N_{\text{AA}} / \text{d} y \text{d} p_{\text{T}}}{\text{d}^{2} \sigma_{p p} / \text{d} y \text{d} p_{\text{T}}},
\end{equation}
\vspace{0.2cm}

\noindent where $\langle N_{coll} \rangle$ corresponds to a scaling factor representing an average number of binary nucleon-nucleon collisions based on the Glauber model predictions \cite{glauber}. $N_{\text{AA}}$ is the average number of J/$\psi$
produced in one minimum-bias A+A collisions and $\sigma_{pp}$ is the cross section in non-single diffractive (NSD) p+p collisions, as a function of transverse momentum $p_{\text{T}}$~and rapidity $y$.
\vspace{1.5cm}

\begin{figure}[H]
\newlength{\extralength}
\setlength{\extralength}{1cm}
\begin{adjustwidth}{-\extralength}{0cm}
 \begin{flushright}
\subfloat[\centering \label{sub:1}]{\includegraphics[width=7.0cm]{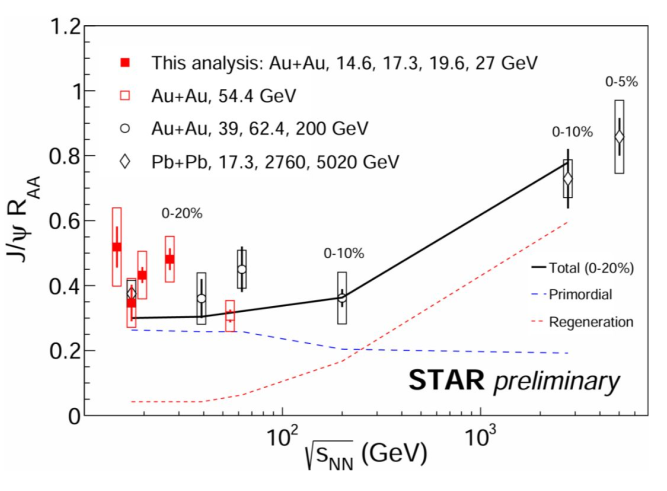}}
\subfloat[\centering \label{sub:2}]{\includegraphics[width=7.0cm]{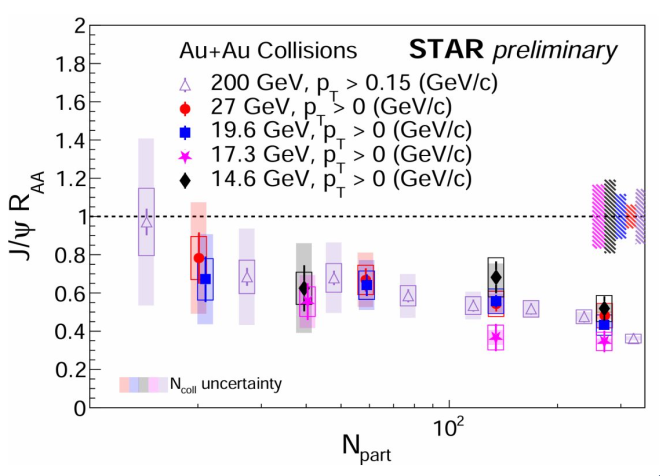}}
 \end{flushright}
\end{adjustwidth}
\caption{(\textbf{a}) J/$\psi$ R$_\text{AA}$ as a function of collision energy with data from the STAR experiment $\sqrt{s_{\mathrm{NN}}} = 14.6 - 200$ GeV in Au+Au collisions \cite{raa1, raa2}, compared with SPS and ALICE measurements \cite{raa3,raa4,raa5} and model predictions \cite{raa6,raa7}. (\textbf{b}) Measurement of the J/$\psi$ R$_\text{AA}$ as a function of centrality in Au+Au collisions for various collision energies at the STAR experiment.\label{fig2}}
\end{figure}

Figure~\ref{sub:1} shows the dependence of the J/$\psi$ R$_\text{AA}$ on collision energy in the range of $\sqrt{s_{\mathrm{NN}}} = 14.6 - 200$ GeV in Au+Au collisions \cite{raa1, raa2}, and includes a comparison with SPS \cite{raa3}, ALICE data \cite{raa4, raa5} and model predictions \cite{raa6, raa7}. Energy dependence can be qualitatively described
by the transport model, with primordial production being dominant at RHIC energies
and regeneration at the LHC, as illustrated by the model curves in the figure.
No significant energy dependence of J/$\psi$ R$_\text{AA}$ is observed in central collisions within uncertainties of up to 200 GeV.

\vspace{0.2cm}

Another measurement, the measurement of J/$\psi$ R$_\text{AA}$ as a function of centrality in Au+Au collisions at different collision energies, is presented in Figure~\ref{sub:2}, where the J/$\psi$ R$_\text{AA}$ indicates a slightly decreasing trend, hence suggesting a stronger suppression in central collisions compared to the peripheral collisions. No significant energy dependence is observed in a given $\langle N_{\text{part}} \rangle$.

\vspace{1cm}

\vspace{1cm}
\newpage
\subsection{$\psi$(2S) over J/$\psi$ Double Ratio}

\begin{figure}[H]
\centering 
\includegraphics[width=9 cm]{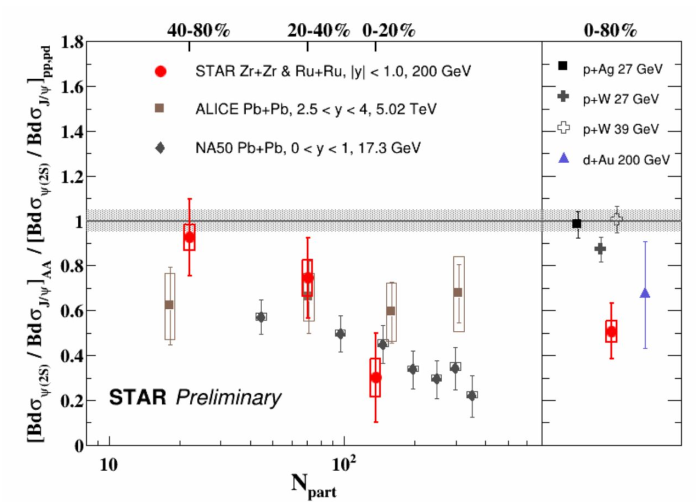}
\caption{$\psi(2 \text{S})$ over $\mathrm{J} / \psi$ double ratio as a function of $N_{\text{part}}$ for various A+A collision systems (left), and the results for $p+$ A systems (right panel). The data are compared with an average $p+p$ reference from NA51, ISR and PHENIX \cite{double1, double2, double3}. \label{figff}}
\end{figure}

The first observation of sequential charmonium suppression in heavy-ion collisions at STAR is
shown in the Figure \ref{figff} and quantified by the $\psi(2 \text{S})$ over $\mathrm{J} / \psi$ double ratio:
\vspace{0.5cm}

\begin{equation}
    \frac{\left[\left(\mathrm{Bd} \sigma_{\psi(2 \text{S})}\right) /\left(\mathrm{Bd} \sigma_{\mathrm{J} / \psi}\right)\right]_{\text{AA}}}{\left[\left(\mathrm{Bd} \sigma_{\psi(2 \text{S})}\right) /\left(\mathrm{Bd} \sigma_{\mathrm{J} / \psi}\right)\right]_{p p, p d}},
\end{equation}
\vspace{0.2cm}

\noindent which is described by the production cross sections of the excited charmonium state $\psi(2 \text{S})$ and of the ground state $\mathrm{J} / \psi$ (i.e. $\sigma_{\psi(2 \text{S})}$ and $\sigma_{\mathrm{J} / \psi}$, respectively) and their branching ratios $B$, comparing heavy-ion (A+A) collisions to proton-proton ($p+p$) or proton-deuteron ($p+d$) collisions.
\vspace{0.2cm}

It is observed that $\psi(2 \text{S})$ is overall more suppressed than $\mathrm{J} / \psi$, reflecting its weaker binding energy and greater sensitivity to the temperature of the QGP. Suppression grows with centrality, driven by higher energy density and QGP lifetime in central collisions.  The data are compared with an average $p+p$ reference from NA51, ISR and PHENIX \cite{double1, double2, double3}. The double ratio appears to be smaller in the isobar system than in the $p+$ A system (see the right panel of the Figure \ref{figff}). Other measurements for different collision systems and energies are further shown for comparison.
\vspace{0.5cm}

\unskip
\vspace{1cm}

\subsection{J/$\psi$ Production vs Multiplicity in $p+p$}
\vspace{0.5cm}

In high multiplicity $p+p$ collisions, multiple parton interactions (MPI) and string percolation effects are expected to play a significant role in modifying the bulk production. These mechanisms can enhance the initial parton density and increase the probability of interactions, potentially affecting the yield of the produced J/$\psi$ mesons.
\vspace{0.2cm}

\begin{figure}[H]
\centering 
\includegraphics[width=8 cm]{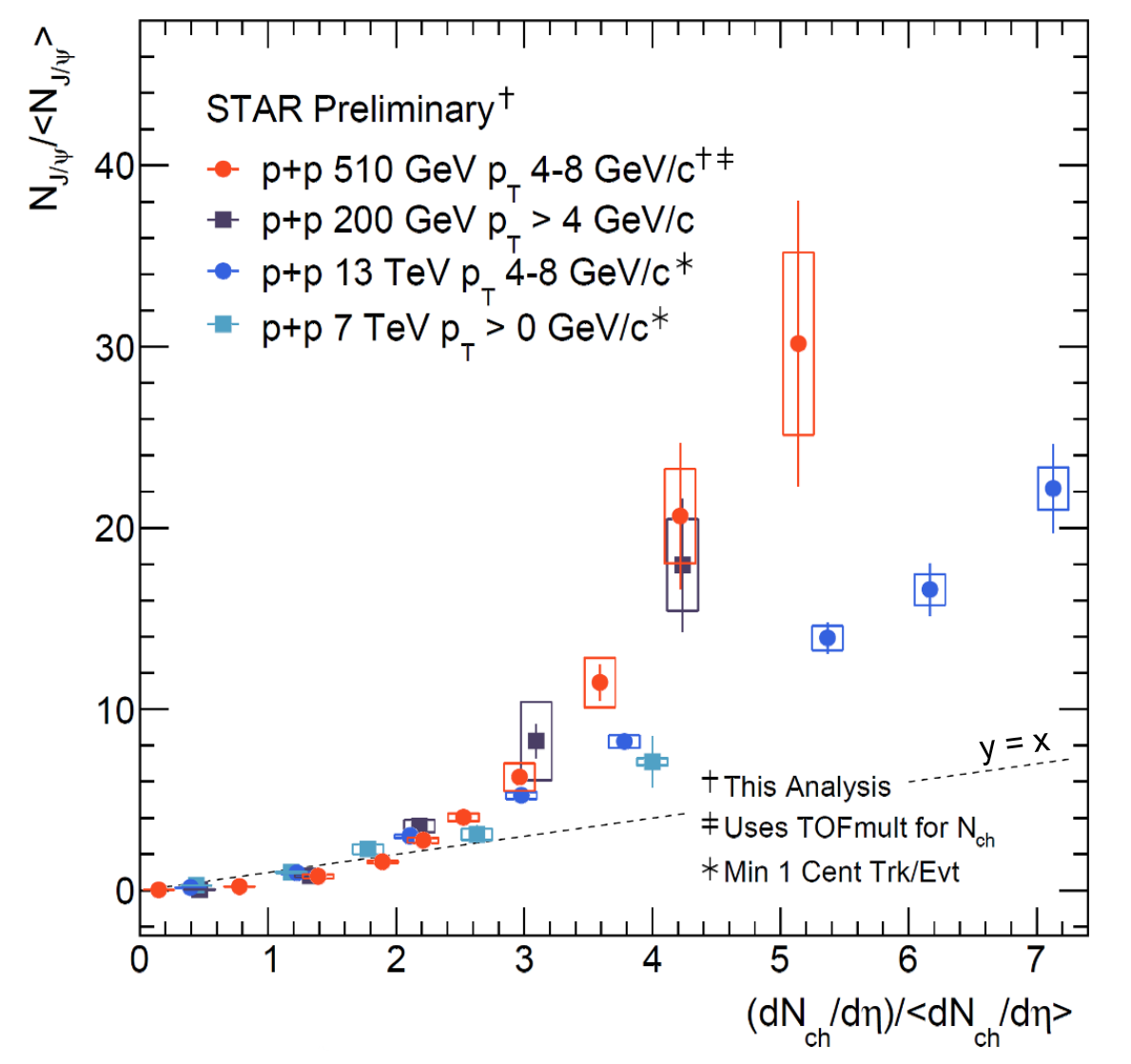}
\caption{The measurement of the J/$\psi$ normalized yields  (relative to their event-average) vs normalized charged particle multiplicity in $p+p$ collisions for different collision energies.\label{fig3}}
\end{figure}

The measured data, showing J/$\psi$ normalized yields vs normalized charged particle multiplicity, are compared to the previous results at $\sqrt{s} = 200$ GeV \cite{pp_multiplicity1} in Figure~\ref{fig3}. The J/$\psi$ normalized yields at $\sqrt{s} = 510$ GeV are consistent with those at $\sqrt{s} = 200$ GeV, with a wider multiplicity range achieved at $\sqrt{s} = 510$ GeV.
There is an indication of a splitting between the results obtained at RHIC and LHC energies \cite{pp_multiplicity2, pp_multiplicity3}. The difference between LHC and RHIC results is still under investigation.
\vspace{0.8cm}

\vspace{1cm}
\newpage
\subsection{J/$\psi$ Production in Jets in $p+p$ collisions}
The study of J /$\psi$ production in jets provides additional discriminative power for production mechanisms. The fraction \textit{z} of charged-particle jet transverse momentum carried by J/$\psi$ is defined as the ratio of the J/$\psi$ transverse momentum to the transverse momentum of the jet:
\vspace{0.2cm}

\begin{equation}
    z(\text{J}/\psi)=\frac{p_{\text{T}}^{\text{J}/\psi}}{p_{\text{T}}^{j e t}}.
\end{equation}
\vspace{0.2cm}

\begin{figure}[H]
\centering 
\includegraphics[width=8 cm]{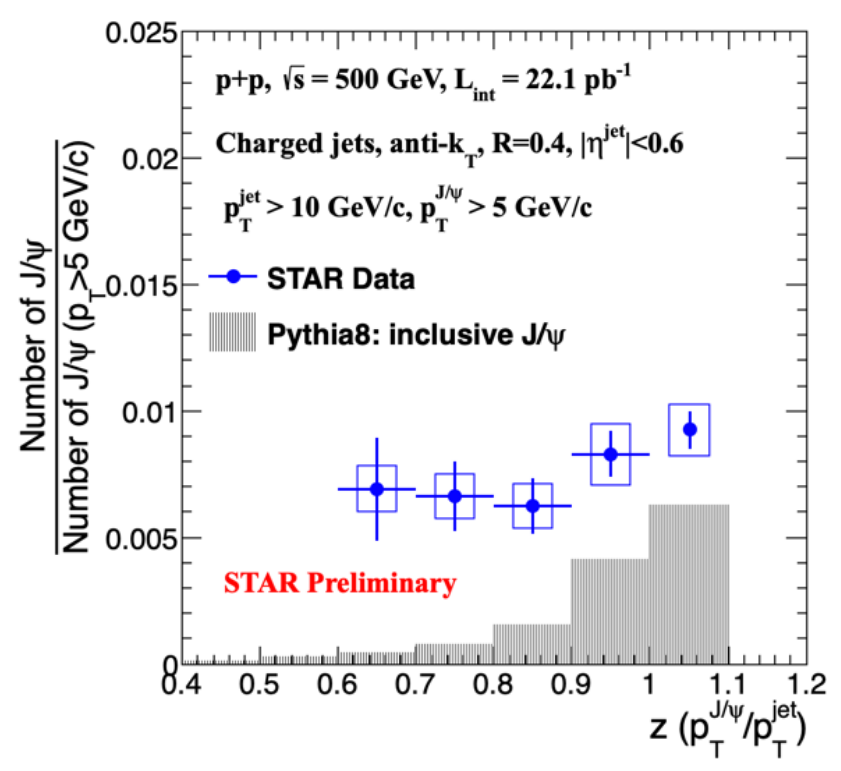}
\caption{The normalized $z$ distributions for inclusive J/$\psi$ mesons produced within a jet in $p+p$ collisions at $\sqrt{s} = 500$ GeV compared to
prediction from PYTHIA 8. The data are normalized
by the J/$\psi$ cross-section at the
same collision energy \cite{jets}.\label{fig4}}
\end{figure}

The measured $z$ distribution for inclusive J/$\psi$ in jets in $p+p$ collisions at $\sqrt{s} $= 500 GeV
normalized by the J/$\psi$ cross-section \cite{jets},
is compared to model prediction (PYTHIA 8) and shown in Figure \ref{fig4}. The results show discrepancy with the model
predictions. The $z$ distribution remains relatively flat,
while PYTHIA predicts a steep rise toward
$z=1$, where most the jet momentum is carried
by the J/$\psi$.

\vspace{0.8cm}

\unskip

\vspace{1cm}

\section{Summary and Outlook}
\vspace{0.2cm}

In these proceedings, we have shown the recent measurements of charmonium production have been shown in A + A collisions,
including the study of J/$\psi$ R$_\text{AA}$  and the charmonium sequential suppression using the $\psi$(2S) over
J/$\psi$ double ratio. The J/$\psi$ production dependence on charged-particle multiplicity at
$\sqrt{s} = 510$ GeV and its production in jets in $p+p$ collisions at $\sqrt{s} = 500$ GeV was
presented.
\vspace{0.2cm}

\begin{figure}[H]
\centering
\includegraphics[width=9 cm]{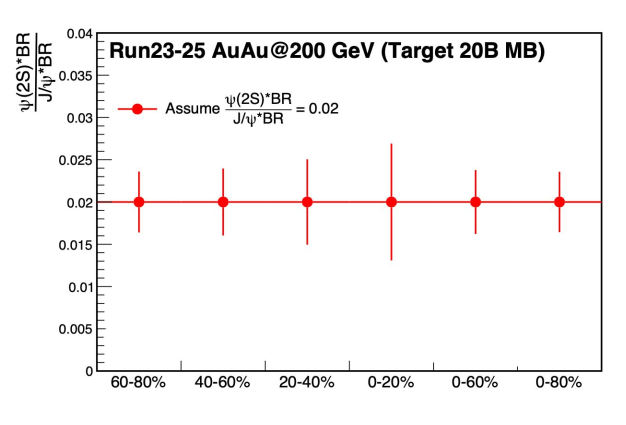}
\caption{Projection of $\psi$(2S) to J$/\psi$ ratios vs centrality for the planned 2023–2025 data-taking period at the STAR experiment \cite{bur}. \label{figpro}}
\end{figure}

Further charmonia measurements have not been covered in this paper, such as the study
of azimuthal anisotropy and polarization.
Studies of J/$\psi$ polarization in jets in $p+p$
collisions are ongoing to provide deeper
insights into the J/$\psi$ production mechanism.
The high luminosity $p+p$ and Au+Au data
at 200 GeV from 2023-2025 \cite{bur} will enable more
precise measurements of J/$\psi$ elliptic anisotropy
and $\psi$(2S) production, as suggested in Figure \ref{figpro}.
\vspace{0.5cm}

\vspace{0.5cm}



\end{document}